\documentclass[10pt,a4paper]{article}
\usepackage[margin=1.5cm]{geometry}
\usepackage[utf8]{inputenc}
\usepackage{listingsutf8,multicol}
\usepackage{amsmath}
\usepackage{subfig}
\usepackage{parskip}
\usepackage{listings}
\usepackage{hyperref}
\usepackage{parcolumns}
\title{Forward-Mode Automatic Differentiation in Julia
       \footnote{This document is an extended abstract that has been accepted for presentation at the AD2016 7th International Conference on Algorithmic Differentiation.}
       \footnote{We thank Kristoffer Carlsson for significant contributions to the library and Isaac Virshup for compiling usage statistics.}}
\author{Jarrett Revels\footnote{Massachusetts Institute of Technology, \texttt{jrevels@csail.mit.edu, mlubin@mit.edu}}, Miles Lubin\footnotemark[2], and Theodore Papamarkou\footnote{University of Glasgow, School of Mathematics and Statistics, \texttt{Theodore.Papamarkou@glasgow.ac.uk}}}
\date{April 2016}

\begin{document}
\maketitle

\section{Introduction}
\label{sec:intro}
\vspace{-0.1cm}

We present ForwardDiff, a Julia package for forward-mode automatic differentiation (AD) featuring performance competitive with low-level languages like C++. Unlike recently developed AD tools in other popular high-level languages such as Python and MATLAB \cite{autograd,algopy,adigator}, ForwardDiff takes advantage of just-in-time (JIT) compilation~\cite{julia} to transparently recompile AD-unaware user code, enabling efficient support for higher-order differentiation and differentiation using custom number types (including complex numbers). For gradient and Jacobian calculations, ForwardDiff provides a variant of vector-forward mode that avoids expensive heap allocation and makes better use of memory bandwidth than traditional vector mode.

In our numerical experiments, we demonstrate that for nontrivially large dimensions, ForwardDiff's gradient computations can be faster than a reverse-mode implementation from the Python-based autograd package. We also illustrate how ForwardDiff is used effectively within JuMP \cite{JuMP}, a modeling language for optimization. According to our usage statistics, 41 unique repositories on GitHub depend on ForwardDiff, with users from diverse fields such as astronomy, optimization, finite element analysis, and statistics.

\vspace{-0.4cm}
\section{Methodology}
\label{sec:methodology}

ForwardDiff implements a Julia representation of a \textit{multidimensional dual number}, whose behavior on scalar functions is defined as:

\begin{equation}\label{eq:univariate}
    f(x + \sum\nolimits_{i=1}^{k} y_i \epsilon_i ) = f(x) + f^{\prime}(x) \sum\nolimits_{i=1}^{k} y_i \epsilon_i,
\end{equation}

where $\epsilon_{i}\epsilon_{j} = 0$ for all indices $i$ and $j$. Storing additional $\epsilon$ components allows for a vector forward-mode implementation of the sort developed by Kahn and Barton \cite{vecmode}. In our formulation, orthogonal $\epsilon$ components are appended to input vector components to track their individual directional derivatives:

\begin{equation}
\vec{x} =
\begin{bmatrix}
x_1 \\
\vdots \\
x_i \\
\vdots \\
x_k
\end{bmatrix} \to
\vec{x}_{\epsilon} =
\begin{bmatrix}
x_1 + \epsilon_1 + 0 \sum_{n=2}^k \epsilon_n \\
\vdots \\
x_i + \epsilon_i + 0 \sum_{n\neq i} \epsilon_n\\
\vdots \\
x_k + \epsilon_k + 0 \sum_{n=1}^{k-1} \epsilon_n
\end{bmatrix}  =
\begin{bmatrix}
x_1 + \epsilon_1 \\
\vdots \\
x_i + \epsilon_i \\
\vdots \\
x_k + \epsilon_k
\end{bmatrix}
\to
f(\vec{x}_{\epsilon}) =
f(\vec{x}) + \sum_{i=1}^k \frac{\partial f(\vec{x})}{\partial x_i} \epsilon_i
\end{equation}

Vector forward mode enables the calculation of entire gradients in a single pass of the program defining $f$, but at the cost of additional memory and operations. Specifically, every dual number must allocate an $\epsilon$ vector of equal size to the input vector, and the number of operations required for derivative propagation scales linearly with the input dimension. In practice, especially in memory-managed languages like Julia, the cost of rapidly allocating and deallocating large $\epsilon$ vectors on the heap can lead to slowdowns that practically outweigh the advantage of fewer passes through $f$.

ForwardDiff's implementation works around this pitfall by stack-allocating the $\epsilon$ vectors, as well as permitting their size to be tunable at runtime relative to the input dimension and performance characteristics of the target function. We call ForwardDiff's strategy \textit{chunk mode}, since it allows us to compute the gradient in bigger or smaller chunks of the input vector. The $\epsilon$ vector length is then the \textit{chunk size} of the computation. For a chunk size $N$ and an input vector of length $k$, it takes $\lceil\frac{k}{N}\rceil$ passes through $f$ to compute $\nabla f(\vec{x})$. For example, it takes two passes through $f$ to evaluate the gradient at a vector of length $k = 4$ and chunk size $N = 2$:

\begin{equation}
\vec{x}_{\epsilon_{1,2}} =
\begin{bmatrix}
x_1 + \epsilon_1 \\
x_2 + \epsilon_2 \\
x_3 \\
x_4
\end{bmatrix} \to
f(\vec{x}_{\epsilon_{1,2}}) = f(\vec{x}) + \frac{\partial f(\vec{x})}{\partial x_1} \epsilon_1 + \frac{\partial f(\vec{x})}{\partial x_2} \epsilon_2
\end{equation}

\begin{equation*}
\vec{x}_{\epsilon_{3,4}} =
\begin{bmatrix}
x_1 \\
x_2 \\
x_3 + \epsilon_1\\
x_4 + \epsilon_2
\end{bmatrix} \to
f(\vec{x}_{\epsilon_{3,4}}) = f(\vec{x}) + \frac{\partial f(\vec{x})}{\partial x_3} \epsilon_1 + \frac{\partial f(\vec{x})}{\partial x_4} \epsilon_2
\end{equation*}

ForwardDiff implements a multidimensional dual number as the type \texttt{Dual\{N,T\}}, where the type parameter \texttt{N} denotes the length of the $\epsilon$ vector and the type parameter \texttt{T} denotes the element type (e.g. \texttt{Dual\{2,Float64\}} has two \texttt{Float64} $\epsilon$ components). This type has two fields: \texttt{value}, which stores the $x$ component, and \texttt{partials}, which stores the stack-allocated $\epsilon$ vector. It's straightforward to overload base Julia methods on the \texttt{Dual} type; here's an example using \texttt{sin}, \texttt{cos}, and \texttt{-} (univariate negation):

\vspace{0.2cm}
\begin{lstlisting}[frame=tb,basicstyle=\small]
import Base: sin, cos, -
sin(d::Dual) = Dual(sin(d.value), cos(d.value) * d.partials)
cos(d::Dual) = Dual(cos(d.value), -(sin(d.value)) * d.partials)
(-)(d::Dual) = Dual(-(d.value), -(d.partials))
\end{lstlisting}

These method definitions are \textit{all} that is required to support the following features:

\begin{itemize}
    \item $n^{th}$-order derivative of \texttt{sin} or \texttt{cos} (through nesting \texttt{Dual} types)
    \item derivative of complex \texttt{sin} or \texttt{cos} via types of the form \texttt{Complex\{Dual\{N,T\}\}}
    \item derivative of \texttt{sin} or \texttt{cos} over custom types, e.g. \texttt{Custom\{Dual\{N,T\}\}} or \texttt{Dual\{N,Custom\}}
\end{itemize}

We unfortunately do not have room in this abstract to adequately cover the latter two items; a proper discussion would require a more thorough exposition of Julia's multiple dispatch and JIT-compilation facilities.

Instead, we discuss how instances of the \texttt{Dual} type can be nested to enable the use of vector-mode AD for higher-order derivatives. For example, the type \texttt{Dual\{M,Dual\{N,T\}\}} can be used to compute \texttt{M x N} $2^{nd}$-order derivatives. As a simple demonstration of the scalar case, we use an instance of the type \texttt{Dual\{1,Dual\{1,Float64\}\}} to take the second derivative of \texttt{sin} at the Julia prompt (The notation \texttt{$\epsilon$[d,k]} is used to denote the $k^{th}$ partial nested at level $d$):

\vspace{0.2cm}
\begin{lstlisting}[frame=tb,basicstyle=\small]
julia> d = Dual(Dual(1.0, 1.0), Dual(1.0, 0.0))
((1.0 + 1.0*$\epsilon$[1,1]) + (1.0 + 0.0*$\epsilon$[1,1])*$\epsilon$[2,1])

julia> d2 = sin(d)
((0.84147 + 0.54030*$\epsilon$[1,1]) + (0.54030 - 0.84147*$\epsilon$[1,1])*$\epsilon$[2,1])

julia> partials(partials(d2, 1), 1)
-0.8414709848078965
\end{lstlisting}

Algebraically, the above example is equivalent to the use of hyper-dual numbers described by Fike and Alonso~\cite{hyperduals}. In fact, a \texttt{Dual} instance with $d$ levels of nesting implements a $d^{th}$-order hyper-dual number, with the added advantage of scaling to arbitrary dimensions. For example, an instance of \texttt{Dual\{M,Dual\{N,Dual\{L,T\}\}\}} can be used to take \texttt{M x N x L} third-order derivatives in a single pass of the target function.

\vspace{-0.4cm}
\section{Performance Analysis}
\label{sec:performance}
In this section, we present timing results for gradient calculations of the Rosenbrock~\eqref{eq:rosenbrock} and Ackley~\eqref{eq:ackley} functions. Recalling~\eqref{eq:univariate} and the discussion in Section~\ref{sec:methodology}, increasing chunk size reduces the number of evaluations of the univariate functions within $f$. We choose Ackley and Rosenbrock as our target functions in order to provide a contrast between the relative gains of increasing chunk sizes when the target function contains many and few expensive univariate functions, respectively.

\begin{equation}\label{eq:rosenbrock}
\text{Rosenbrock}(\vec{x}) = \sum\nolimits_{i=1}^{k-1} 100 (x_{i+1} - x_{i}^2)^2 + (1 - x_i)^2
\end{equation}

\begin{equation}\label{eq:ackley}
\text{Ackley}(\vec{x}) = -a \exp\left( -b \sqrt{\frac{1}{k} \sum\nolimits_{i=1}^k x^{2}_{i}} \right) - \exp\left(\frac{1}{k} \sum\nolimits_{i=1}^k \cos(cx_{i})\right) + a + \exp(1)
\end{equation}

Table~\ref{tab:cpp} shows evaluation times for calculating gradients of our two target functions using ForwardDiff versus a naive equivalent C++ implementation. Various chunk sizes were tested, while the input size was fixed at $12000$ elements. For the sake of simplicity, ForwardDiff's \texttt{Dual\{N,T\}} type was translated into a hardcoded C++ class for each $N$.

\vspace{-0.3cm}
\begin{table}[htb]
\begin{center}
\subfloat[$\nabla(\text{Rosenbrock})$]{\begin{tabular}{lll}
chunk size $N$ & C++ Time (s) & ForwardDiff Time (s)\\ \hline
1              & 2.66744      & 0.62760 \\
2              & 2.71184      & 0.45541 \\
3              & 1.92713      & 0.44469 \\
4              & 1.45306      & 0.42354 \\
5              & 1.24949      & 0.44045
\end{tabular}}
\hfil
\subfloat[$\nabla(\text{Ackley})$]{\begin{tabular}{lll}
chunk size $N$ & C++ Time (s) & ForwardDiff Time (s)\\ \hline
1              & 4.02078      & 5.12890 \\
2              & 4.35398      & 2.72003 \\
3              & 3.05532      & 1.86055 \\
4              & 2.26095      & 1.47578 \\
5              & 1.91985      & 1.23500
\end{tabular}}
\end{center}
\caption{Time to evaluate gradients using C++ vs. ForwardDiff, input size $k = 12000$}
\label{tab:cpp}
\end{table}

Table~\ref{tab:cpp} helps illustrate that the optimal chunk size for a given problem is a result of a trade-off between memory bandwidth, memory alignment, cache performance, and function evaluation cost. For example, note that ForwardDiff's $\nabla(\text{Rosenbrock})$ performance worsens when going from $N=4$ to $N=5$, and that the C++ implementation's performance with both functions worsens when going from $N=1$ to $N=2$. The former observation is likely due to the large memory bandwidth cost relative to the cost of the arithmetic operations, while the latter observation is likely due to poor alignment of input vector (since each 2-dimensional dual number is essentially a struct of three \texttt{double} values - one for the instance value, and two for the partial components).

\begin{table}[htb]
\begin{center}
\begin{tabular}{lllll}
Function        &  Input Size $k$  & autograd Time (s) & ForwardDiff Time (s) & ForwardDiff Multithreaded Time (s)\\ \hline
Ackley          &  10              & 0.001204          & 0.000007             & 0.000007 \\
Ackley          &  100             & 0.008472          & 0.000058             & 0.000056 \\
Ackley          &  1000            & 0.081499          & 0.006351             & 0.002620 \\
Ackley          &  10000           & 0.835441          & 0.564828             & 0.253798 \\
Ackley          &  100000          & 8.361769          & 56.850198            & 24.394373 \\
Rosenbrock      &  10              & 0.000866          & 0.000003             & 0.000003 \\
Rosenbrock      &  100             & 0.004395          & 0.000034             & 0.000041 \\
Rosenbrock      &  1000            & 0.040702          & 0.003010             & 0.001582 \\
Rosenbrock      &  10000           & 0.411095          & 0.302277             & 0.159703 \\
Rosenbrock      &  100000          & 4.173851          & 30.365882            & 14.111776
\end{tabular}
\end{center}
\caption{Time to evaluate gradients using autograd (reverse mode) vs. ForwardDiff, chunk size $N = 10$}
\label{tab:autograd}
\end{table}

Table~\ref{tab:autograd} compares the gradient computation time of the \textit{reverse-mode} implementation of the Python-based autograd package versus the forward-mode implementation of ForwardDiff for varying input sizes. We also include results obtained using our experimental multithreaded implementation, which show a $\sim$2x speed-up using 4 threads compared to our single-threaded implementation.

Both functions have linear complexity in the input dimension $k$; therefore reverse mode, which requires $O(1)$ passes through each function, scales linearly, while our forward mode, which requires $O(k)$ passes through each function (with fixed $N$), scales quadratically. The results in Table~\ref{tab:autograd} agree with this complexity analysis. Nevertheless, there is a huge performance gap between these two implementations such that autograd is slower on these examples when $k \le 10000$, despite reverse mode being a superior algorithm in principle for computing gradients.

The code used to generate the timings in this section can be found at \url{https://github.com/JuliaDiff/ForwardDiff.jl/tree/jr/benchmarks/benchmark}. Julia benchmarks were run using Julia version 0.5.0-dev+3200, C++ benchmarks were compiled with clang-600.0.57 using \texttt{-O2}, and Python benchmarks were run using Python version 2.7.9.

\vspace{-0.5cm}
\section{ForwardDiff within JuMP}

Effective use of ForwardDiff has brought improvements to JuMP~\cite{JuMP}, a domain-specific language for optimization embedded in Julia where users provide closed-form algebraic expressions using a specialized syntax. JuMP, and similar commercial tools like AMPL \cite{AMPLbook}, compute derivatives of user models as input to nonlinear optimization solvers, which is quite different from ForwardDiff's original target case of differentiating general user-defined code.

JuMP computes sparse Hessians by using the graph coloring approach of~\cite{GebIjoc}, which requires computing a small number of Hessian-vector products in order to recover the full Hessian. JuMP's forward-over-reverse mode implementation for Hessian-vector products makes use of ForwardDiff's chunk mode, essentially computing Hessian-matrix products instead of independent Hessian-vector products. This use of chunk mode yielded speedups of 30\% on benchmarks presented in \cite{JuMP} (under review). The results in \cite{JuMP} include this speedup but are not accompanied by a discussion of the methodology of chunk mode.

On the user-facing side, ForwardDiff has enabled JuMP to be the first AML to our knowledge which performs automatic differentiation of user-defined functions embedded within closed-form expressions. We reproduce an example from~\cite{JuMP} illustrating a user-defined square root function within a JuMP optimization model:

\noindent\makebox[\linewidth]{\rule{\textwidth}{0.4pt}}
\vspace{-0.7cm}
\begin{lstlisting}[multicols=2,basicstyle=\small]
function squareroot(x)
    # Start Newton's method at x
    z = x
    while abs(z*z - x) > 1e-13
        z = z - (z*z-x)/(2z)
    end
    return z
end
JuMP.register(:squareroot, 1,
               squareroot, autodiff=true)
m = Model()
@variable(m, x[1:2], start=0.5)
@objective(m, Max, sum(x))
@NLconstraint(m,
     squareroot(x[1]^2+x[2]^2) <= 1)
solve(m)
\end{lstlisting}
\vspace{-0.7cm}
\noindent\makebox[\linewidth]{\rule{\textwidth}{0.4pt}}

JuMP computes gradients of \texttt{squareroot} with ForwardDiff which are then integrated within the reverse-mode computations of JuMP. We do not yet support $2^{nd}$-order derivatives of user-defined functions. While this functionality is immature and leaves room for improvement (we could attempt to tape the user-defined functions and calculate their gradients in reverse mode), it already creates a new and useful way for JuMP users to seamlessly interact with AD when small parts of their model cannot easily be expressed in closed algebraic form.

\label{sec:jump}

\vspace{-0.5cm}
\section{Future Work}

We are currently investigating several avenues of research that could improve ForwardDiff's performance and usability. We are in the preliminary phases of implementing SIMD vectorization of derivative propagation. We intend to address perturbation confusion \cite{pertubationconfusion} by intercepting unwanted pertubations at compile time using Julia's metaprogramming capabilities. Finally, we wish to improve our support for matrix operations such as eigenvalue computations by directly overloading linear algebra functions, a technique which has already seen use in \cite{algopy}.

\vspace{-0.5cm}
\bibliographystyle{unsrt}
\bibliography{main}

\end{document}